# Flux growth at ambient pressure of millimeter-sized single crystals of LaFeAsO, $LaFeAsO_{1-x}F_x$, and $LaFe_{1-x}Co_xAsO$

J.-Q. Yan[1], S. Nandi[1,2], J. L. Zarestky[1], W. Tian[1], A. Kreyssig[1,2], B. Jensen[1], A. Kracher[1], K. W. Dennis[1], R. J. McQueeney[1,2], A. I. Goldman[1,2], R. W. McCallum[1,3], and T. A. Lograsso[1]

[1] Division of Materials Science and Engineering, Ames Laboratory, US-DOE, Iowa State University, Ames, Iowa 50011, USA

[2] Department of Physics and Astronomy, Iowa State University, Ames, Iowa 50011, USA

[3] Department of Materials Science and Engineering, Iowa State University, Ames, Iowa 50011, USA

Millimeter-sized single crystals of LaFeAsO, $LaFeAsO_{1-x}F_x$, and $LaFe_{1-x}Co_xAsO$ were grown in NaAs flux at ambient pressure. The detailed growth procedure and crystal characterizations are reported. The as-grown crystals have typical dimensions of 3 × 4 × 0.05-0.3 mm$^3$ with the crystallographic *c*-axis perpendicular to the plane of the plate-like single crystals. Some crystals manifest linear dimensions as large as 4-5 mm. X-ray and neutron single crystal scattering confirmed that LaFeAsO crystals exhibit a structural phase transition at $T_s \sim$ 154 K and a magnetic phase transition at $T_{SDW} \sim$ 140 K. The transition temperatures agree with those determined by anisotropic magnetization, in-plane electrical resistivity and specific heat measurements and are consistent with previous reports on polycrystalline samples. Co and F were successfully introduced into the lattice leading to superconducting $LaFe_{1-x}Co_xAsO$ and $LaFeAsO_{1-x}F_x$ single crystals, respectively. This growth protocol has been successfully employed to grow single crystals of NdFeAsO. Thus it is expected to be broadly applicable to grow other

*RM*AsO (*R* = rare earth, *M* = transition metal) compounds. These large crystals will facilitate the efforts of unraveling the underlying physics of iron pniticide superconductors.

After the initial report of superconductivity with $T_c$ ~ 26 K in $LaFeAsO_{1-x}F_x$,[1] the maximum superconducting transition temperature for this class of materials was quickly raised to ~ 55 K by replacing La with other rare earth elements[2] or applying an external pressure.[3] Shortly after, superconductivity with $T_c$ up to 38 K was observed in doped $AFe_2As_2$ (*A* = Ca, Sr, Ba, and Eu, "122") compounds,[4] which share the same structural unit of FeAs layers with the *R*FeAsO (*R* = rare earth, "1111") system. Although the 1111 system was discovered earlier and manifests a higher $T_c$, the focus of the scientific community has shifted toward the 122 systems because (1) sizeable high quality single crystals of the 122 system have been successfully produced by many groups and both electron and hole doping can be systematically manipulated,[5,6,7,8,9,10,11,12,13,14] and (2) the growth of large single crystals of the 1111 system has been proven to be difficult so that the largest crystals are still in submillimeter size despite tremendous efforts.[15,16,17] Under ambient pressure, plate-like $NdFeAsO_{1-x}F_x$ crystals with the lateral size ranging from 5 to 30 μm and thickness of 1 - 5 μm can be grown by the flux method using NaCl as the flux.[17] The growth takes 10 days to homogenize at 1050 °C which suggests the solubility of NdFeAsO in NaCl flux is quite limited at this temperature. Growths at even higher temperature appear necessary but may well be limited by the stability of *R*FeAsO compounds and/or the high vapor pressure of As. High pressure growth has generally been more successful in growing larger crystals from the NaCl flux and typical crystals have dimensions on the order of 300 μm.[15] A recent paper reported the growth of CeFeAsO single crystals with the largest dimension reaching 600 μm in Sn-flux,[18] to the best of our knowledge, the largest 1111 single crystal reported to date.

We have found that NaAs is an effective solvent for *R*FeAsO and consequently we have succeeded in growing a high yield of sizeable high quality single crystals by the flux method under ambient pressure. In this Letter, we report the details of the flux growth and characterization of both undoped and doped LaFeAsO single crystals.

Both prefired LaFeAsO and precursor mixture of LaAs, 1/3 $Fe_2O_3$, and 1/3 Fe were used as the charge in the growth. LaFeAsO was prepared by conventional solid state reaction method.[1] LaAs was synthesized by reacting pure La filings and As pieces in a quartz ampoule partially filled with Ar. The ampoule was ramped to 600 °C at the rate of 50 °C per hour, held at temperature for 15 hours, and then heated to 900 °C in 3 hours and held 15 hours. NaAs was prepared by reacting Na and As chunks in a sealed Ta tube. The Ta tube was sealed in a quartz ampoule and heated to 600 °C at the rate of 30 °C per hour and held at temperature for 12 hours. For safety, all reactions which involved elemental As precursor materials were carried out with the sealed ampoules placed in a superalloy retort flushed with flowing nitrogen which was discharged into the laboratory fume exhaust system. The charge and NaAs flux with the molar ratio of 1: 20 were mixed and sealed in a Ta tube under ~1/3 atmosphere of argon gas. The Ta tube was then sealed in an evacuated quartz tube. To ensure safety, the quartz tube was then loaded into an Inconel tube. The entire assembly was then heated in a tube furnace, with a quartz retort under flowing nitrogen, to 1150 °C at a rate of 90 °C per hour. After holding at 1150 °C for 24 hours, it was cooled to 600 °C at 3 °C per hour followed by a furnace quench to room temperature. As NaAs is highly hydrophilic, the separation of the crystals from flux is easily accomplished by rinsing them with deionized water in a closed fume hood.

The inset of Fig. 1(a) shows a picture of a single crystal of LaFeAsO against a mm scale. The as-grown crystals are plate-like with typical dimensions of $3 \times 4 \times 0.05\text{-}0.3$ mm$^3$. Some crystals manifest linear dimensions as large as 4-5 mm. The crystallographic *c*-axis is perpendicular to the plane of the plate-like single crystals. Figure 1(a) shows a rocking curve for the (1 1 7) reflection. The high resolution x-ray diffraction measurements were performed on a crystal mounted in a standard four-circle diffractometer using Cu $K\alpha$ radiation selected by a Ge (111) monochromator from a rotating anode X-ray source. The measured mosaicity of the crystal was 0.04° full width half maximum for the (1 1 7) reflection at room temperature, indicating the excellent quality of the single crystal. Room temperature powder x-ray diffraction on crushed crystals confirmed that the crystals are single phase (see Fig. 1(b)) with lattice parameters, $a$ = 4.022(2) Å and $c$ = 8.746(6) Å, consistent with previous reports.[19,20,21] Elemental analysis performed using wavelength dispersive x-ray spectroscopy (WDS) in a JEOL JXA-8200 Superprobe electron probe microanalyzer (EPMA) confirmed the atomic ratio of 1:1:1:1 with no noticeable compositional variation of the atomic ratio across the crystal. In particular, we carefully sampled for evidence of Na on clean and cleaved surfaces of various pieces and none was found. The measured structural and magnetic transition temperatures, presented below, agree with previous reports on polycrystalline samples, ruling out the possible substitution of La by Na. Since the starting materials were also exposed to quartz or Ta tubes, we also searched for possible incorporation of Si and Ta. No sign of Si was found, but we observed ~0.2% (atomic)

Ta homogeneously distributed in the bulk. To avoid Ta incorporation, attempts were made to prepare NaAs in $Al_2O_3$ tubes and grow crystals in $Al_2O_3$ crucibles. The reaction of NaAs flux with the $Al_2O_3$ crucibles resulted in failure of this approach.

Single crystals of comparable quality could be obtained by starting with either prefired polycrystalline LaFeAsO or precursor mixture (LaAs + 1/3 $Fe_2O_3$ + 1/3 Fe). The former generally leads to crystals with smaller lateral size. Also, depending on the purity of starting materials, needle like FeAs crystals were sometimes observed.

The structural ($T_s$ = 153 K) and magnetic ($T_{SDW}$ = 140 K) transitions of LaFeAsO are accompanied by anomalies in the temperature dependence of magnetization, electrical resistivity, and specific heat. Figure 2 (a) shows the temperature dependence of magnetic susceptibility $\chi = M(T)/H$ measured with a Quantum Design (QD) Magnetic Property Measurement System (MPMS). There is a clear anisotropy with $\chi_{ab} > \chi_c$ over the entire temperature range of 2 to 300 K. While cooling, both $\chi_{ab}$ and $\chi_c$ show a step-like anomaly starting at $T_s \sim 153$ K. The starting temperature and the Curie-Weiss-like tail at low temperatures agree well with previous reports on magnetic properties of polycrystalline samples.[21,22]

Six separate single crystals from the same batch were selected for measurements of the in-plane electrical resistivity using a QD Physical Property Measurement System (PPMS). Electrical contact was made to the samples using LakeShore silver epoxy to attach Cu wires in a 4-probe configuration. Two typical behaviors, designated R1 and R2, were observed as shown in Fig. 2(b). The decrease in resistivity below $T_s$ for R1 is similar to that found for polycrystalline samples. In contrast, R2 increases while cooling below $T_s$. A careful WDS study of these six crystals did not find any noticeable difference of Ta-content. Therefore, it is unlikely that the discrepancy is due to the Ta incorporation. On the other hand, we found that all crystals grown from a mixture with the charge : flux molar ratio of 1 : 10 shows the same R2-type temperature dependence. This indicates that the resistivity anomalies may be sensitive to defects formed in the growth process since the higher dilution ratio generally results in lower growth rates for the same cooling rate. From the temperature dependence of d[R(T)/R(300)]/dT of R1, the structural and magnetic transitions are identified at $T_s$ = 153 K and $T_{SDW}$ = 140 K, respectively.

Figure 2(c) presents the temperature dependent specific heat measured using a QD PPMS. One lambda-type anomaly is clearly observed at approximately 150 K. The inset highlights the details of the anomaly after subtraction of a smooth background determined by fitting the specific heat data well below and above the lambda anomaly by a polynomial function. The structural and magnetic transition temperatures are determined to be 157 K and 136 K, respectively, following Kondrat *et al.*[22]

To confirm the structural and magnetic transitions in our crystal, we performed temperature dependent, single crystal x-ray and neutron diffraction experiments. Temperature-dependent, single crystal x-ray diffraction measurements were performed using high energy x-rays at the station 6-ID-D in the MU-CAT sector of the Advanced Photon Source, Argonne National Laboratory. The energy of incident x-ray energy was 129 keV, ensuring full penetration of the sample with dimensions of 2×2×0.2 $mm^3$. The beam size was 0.2×0.2 $mm^2$. To record the two dimensional diffraction patterns, a MAR345 image plate detector was positioned 2192.5 mm behind the sample. The sample, with the ***c***-axis parallel to the incident beam, was attached to Kapton tape and mounted on a closed cycle, Displex refrigerator. The sample was rocked by ± 3.2° about both axes perpendicular to the incoming beam. Rocking the sample through these small angular ranges provides an image of reciprocal space planes that lie normal to the beam direction.[23] The diffraction patterns were recorded while the temperature was varied between 7 and 165 K. Figure 3 (a) summarizes the single crystal diffraction results. Above $T_s$, high energy x-ray diffraction revealed only diffraction spots corresponding to the tetragonal space group *P*4/*nmm* (see right panel of Fig. 3(b)), consistent with the earlier report.[1] Below $T_s$ = 154.5 K, however, the spots split (see left panel of Fig. 3(b)) according to the twinning law described in Ref. [24], consistent with a tetragonal (*P*4/*nmm*)-to-orthorhombic (*Cmma*) phase transition. The small splitting (≈0.0024) observed at 120 K is, again, consistent with the previous report on polycrystalline samples.[20] The evolution of orthorhombic distortion as a function of temperature was extracted from the splitting of the (2 2 0) diffraction spots using the program FIT2D[25] and shown in Fig. 3(a). The sharp increase of the orthorhombic distortion below $T_s$ ~ 154.5 K confirms the structural transition from tetragonal *P*4/*nmm* to orthorhombic *Cmma* in our LaFeAsO crystals.

Single crystal neutron diffraction was performed on a "large" crystal with the mass of ~ 5 mg. Magnetization measurements on this single crystal confirmed that the magnetic anomaly occurs at the same temperature as the crystal used in the x-ray diffraction study. Single crystal neutron diffraction measurement was performed using the Ames Laboratory HB1A fixed-incident-energy triple-axis spectrometer at the High Flux Isotope Reactor of Oak Ridge National Laboratory. Collimations of 48'-48'-sample-40'-68' downstream from reactor to detector were used. The single crystal was oriented in the (hhl) scattering plane, sealed in a helium exchange gas sample can, and cooled to a base temperature of ~10 K using a closed-cycle helium refrigerator. The order parameter was measured by sitting on the (1/2 1/2 3/2) magnetic peak position and slowly ramping the temperature from 10 K to 180 K while counting in ~ 3 minutes time intervals. The temperature dependence of the intensity of the magnetic reflection(1/2 1/2 3/2) is also presented in Fig. 3(a) to highlight the separation of the structural and magnetic phase transitions. The result clearly shows the onset of magnetic order at $T_{SDW}$ ~ 140 K, which is about 14 K below $T_s$.

Having demonstrated that NaAs is a suitable flux for the growth of the parent compound LaFeAsO, we extended the growth to the doped $LaFe_{1-x}Co_xAsO$ and $LaFeAsO_{1-x}F_x$ series by partially replacing Fe with Co and NaAs with NaF, respectively. A number of single crystals with various doping contents have successfully been grown.WDS analysis of cleaved surfaces confirms that the doping element entered into the lattice (see Fig. 4 (a) and (b)) for both series. As with the parent compound, WDS analysis found no sign of Na but ~0.2% (atomic) of Ta in all of our doped crystals. It's noteworthy that NaF might not be a good source for introducing F. With NaF in flux, the yield of crystals growth was normally low and the crystals were smaller. Figures 4 (c) and (d) show the low-field magnetization and in-plane resistivity of two superconducting compositions. While the extent of Co doping could be accurately determined by WDS analysis, WDS has difficulty in accurately determining the ratio of O : F at these levels. For the crystal shown in Fig. 4(a), a tentative composition suggested by WDS is $LaFeAsO_{0.91}F_{0.09}$, while superconductivity was observed at $T_c$ ~ 10 K and 12 K from magnetization and electrical resistivity measurements, respectively. According to the phase diagram of $LaFeAsO_{1-x}F_x$,[1] a $T_c$ ~ 12 K suggests a composition of $LaFeAsO_{0.97}F_{0.03}$ with the F-content lower than our estimate. Whether this discrepancy arises from the uncertainty of our elemental analysis, or the suppression of superconductivity by 0.2% (atomic) Ta, remains to be resolved.

In summary, we have successfully grown LaFeAsO, $LaFe_{1-x}Co_xAsO$, and $LaFeAsO_{1-x}F_x$ crystals with sizes up to 4-5 mm from a NaAs flux at ambient pressure. Undoped LaFeAsO crystals show the same structural and magnetic transition temperatures as reported for polycrystalline samples. Detailed characterization of $LaFe_{1-x}Co_xAsO$ and $LaFeAsO_{1-x}F_x$ compositions are in progress to map out the phase diagrams. Meanwhile, proof of concept growths of doped and undoped NdFeAsO crystals have successfully been performed and we therefore expect that NaAs flux can be applied to crystal growth of other *RM*AsO (*M* = transition metal) members. The availability of these sizeable single crystals will enable comparisons to be made with $AFe_2As_2$ systems and cuprate superconductors using a variety of techniques, hopefully leading to a complete understanding of the iron pnictide superconductors.

**Acknowledgements**

JQY thanks Prof. P. C. Canfield for fruitful discussions and for making part of the synthesis possible. The assistance of D. S. Robinson in performing the HEXRD studies at the APS is highly appreciated. SN thanks M. G. Kim for crystal orientation. Ames Laboratory is operated for the US Department of Energy by Iowa State University under Contract No. DE-AC02-07CH11358. Use of the Advanced Photon Source was supported by US DOE under Contract No. DE-AC02-06CH11357. The HFIR Center for Neutron Scattering is a national user facility funded by the United States Department of Energy, Office of Basic Energy Sciences, Materials Science, under Contract No. DE-AC05-00OR22725 with UT-Battelle.

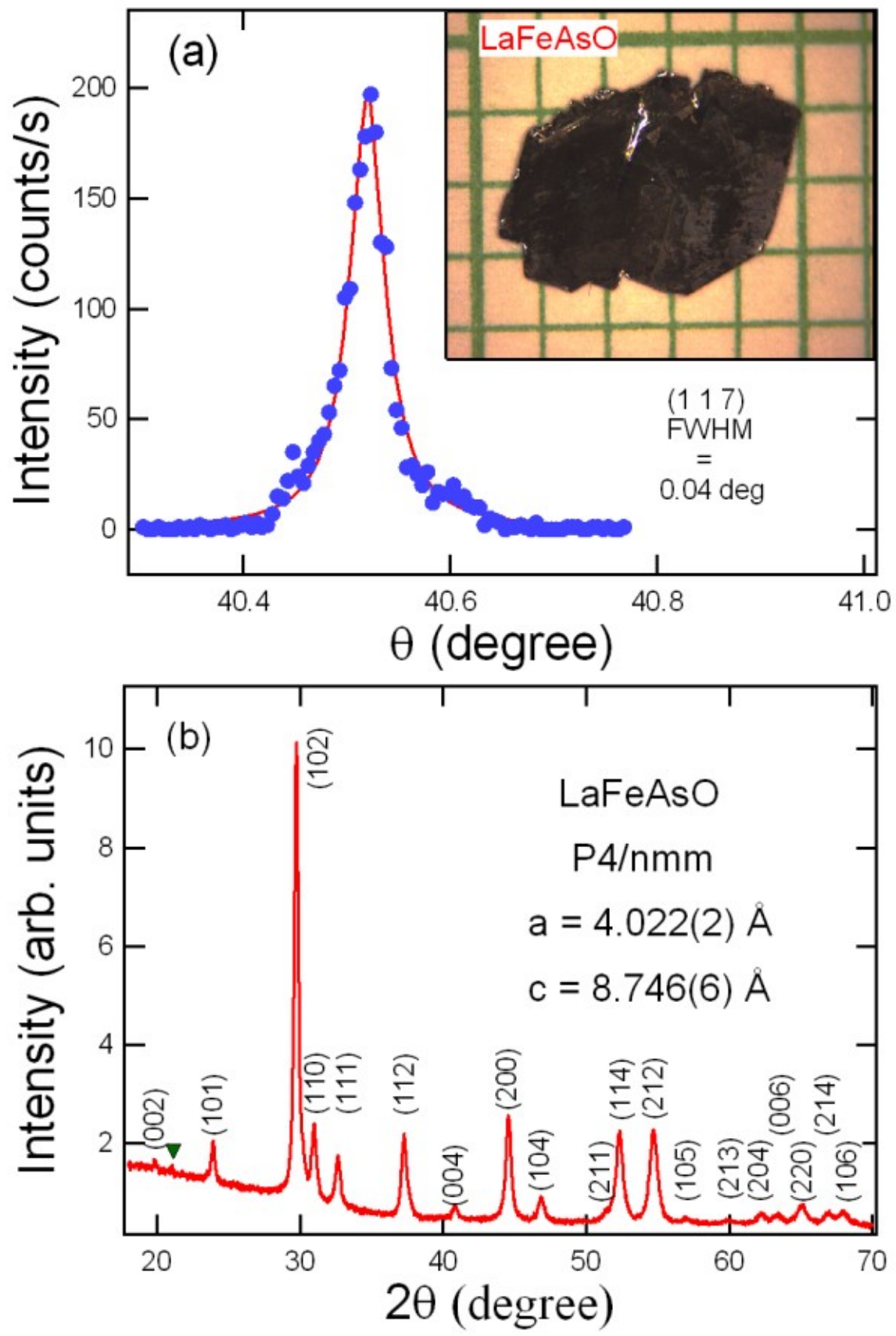

Figure 1 (color online) Room temperature x-ray characterization of LaFeAsO crystals. (a) Rocking curve through the (1 1 7) reflection of the LaFeAsO single crystal used for the x-ray diffraction study. Inset: picture of a LaFeAsO single crystal on millimeter grid paper. (b) Powder x-ray diffraction pattern of pulverized LaFeAsO single crystals. The weak extra diffraction peak (▼) at $2\theta \sim 21^{o}$ is from sodium arsenate hydrate.

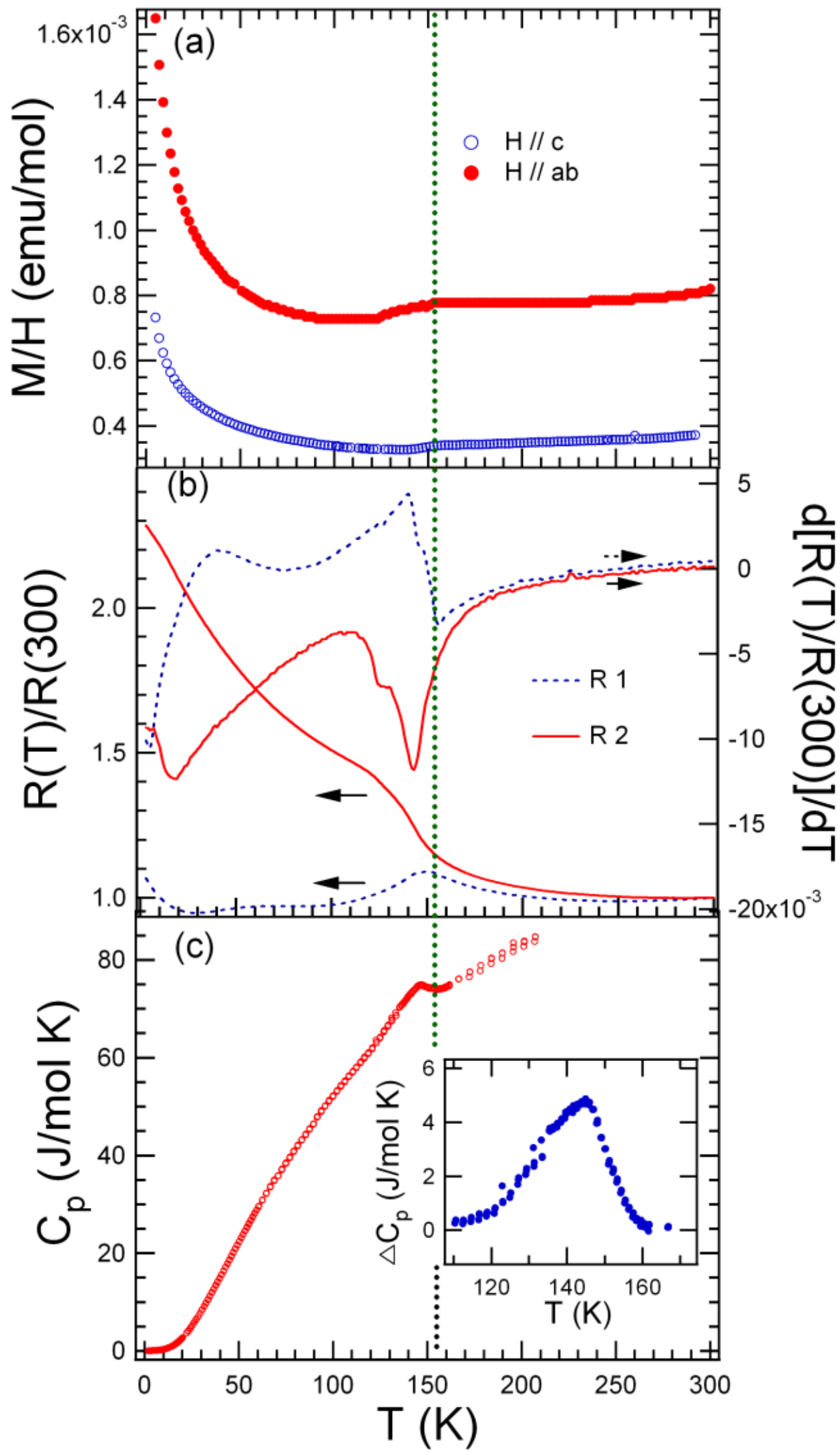

Figure 2 (color online) Physical properties of grown LaFeAsO crystals. (a) Temperature dependence of magnetization measured under a magnetic field of 30 kOe. (b) Examples of the two types of temperature dependent in-plane resistivity. (c) Temperature dependence of specific heat. Inset shows the background subtracted data. The vertical dashed line is a guide for eyes.

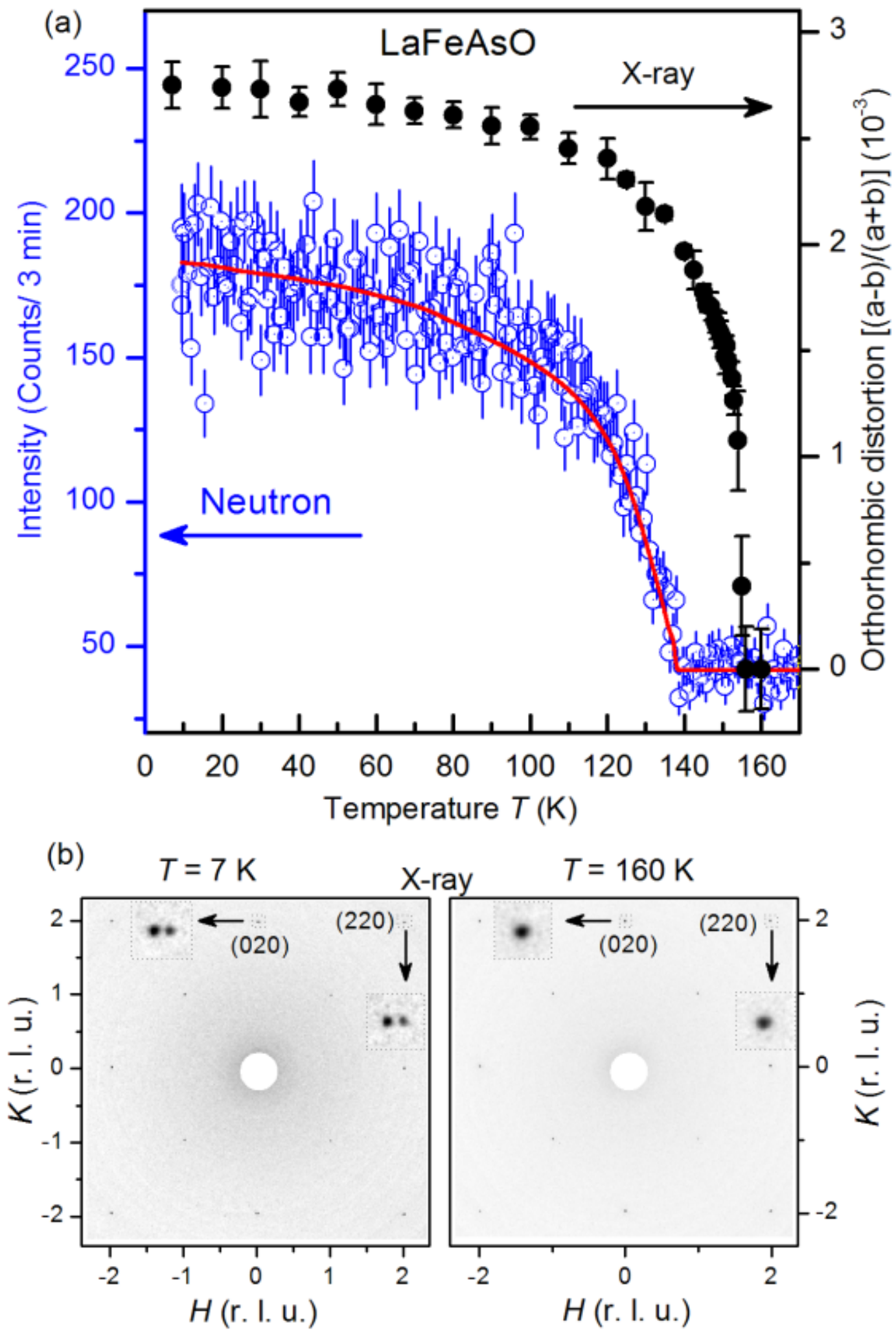

Figure 3 (color online) Temperature dependent single crystal x-ray and neutron diffraction study of LaFeAsO crystals. (a) The temperature dependence of orthorhombic distortion and intensity of the magnetic reflection (1/2 1/2 3/2). (b) High energy x-ray diffraction patterns of the single crystal at T = 7 K and 160 K, respectively. The insets display the (220) and (020) peaks in greater detail.

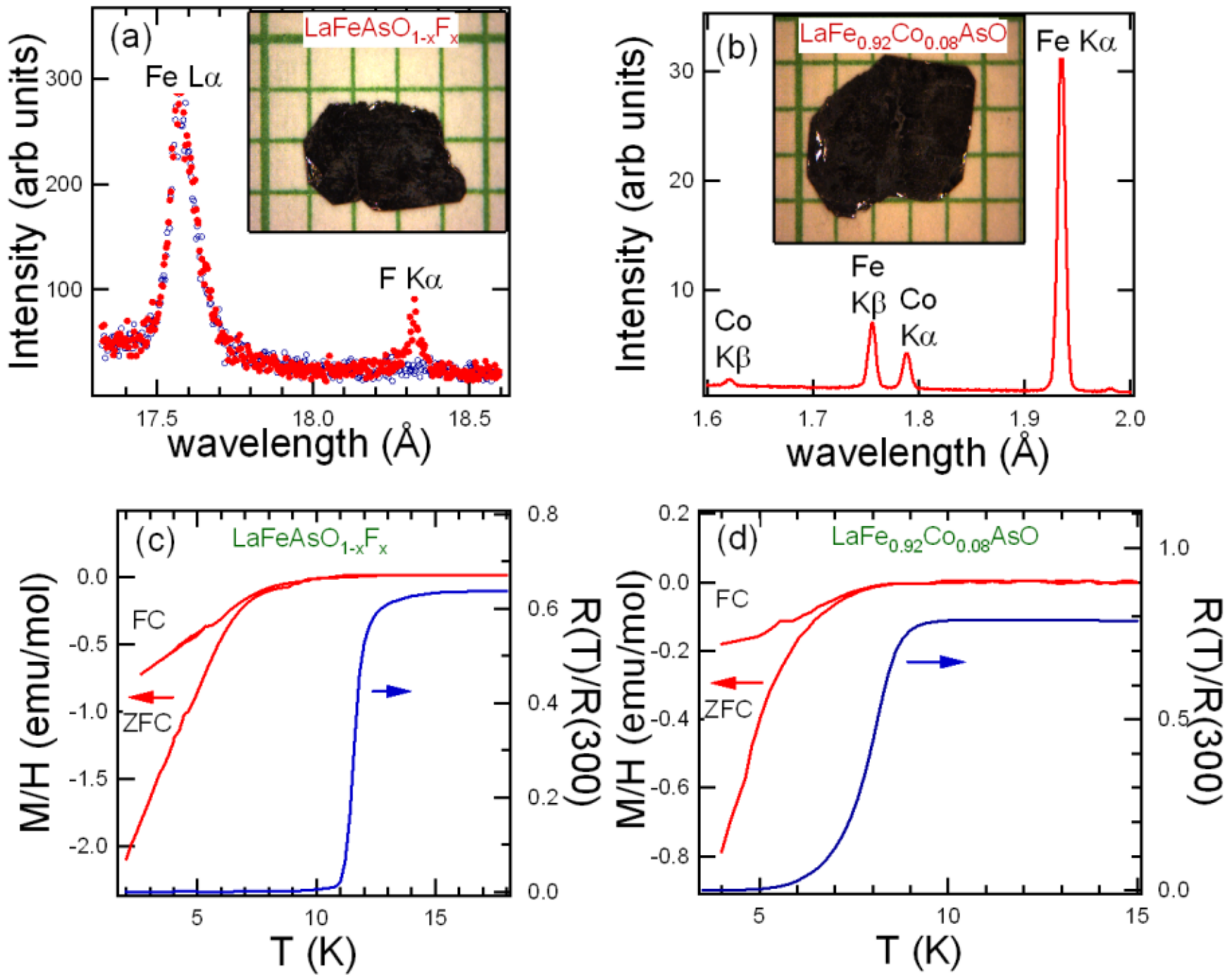

Figure 4 (color online) Characterization of superconducting compositions. (a) Electron microprobe spectrum of x-ray intensity versus wavelength in the vicinity of the F $K_{\alpha}$ line (18.32 Å) for the F-doped sample (●) compared to a F-free sample (○). Inset shows a picture of a $LaFeAsO_{1-x}F_x$ single crystal on millimeter grid paper. (b) Electron microprobe spectrum of x-ray intensity versus wavelength for Co-doped LaFeAsO. Inset shows a picture of a $LaFe_{0.92}Co_{0.08}AsO$ single crystal on millimeter grid paper. (c) Low-field M/H data measured under a field of 35 Oe perpendicular to *c*-axis and in-plane electrical resistivity for a F-doped crystal. (d) Low-field M/H data measured under a field of 35 Oe perpendicular to *c*-axis and in-plane electrical resistivity for $LaFe_{0.92}Co_{0.08}AsO$ crystal.

**References**

[1] Y. Kamihara, T. Watanabe, M. Hirano, and H. Hosono, J. Am. Chem. Soc. **130,** 3296 (2008).

[2] Z. A. Ren et al., Chin. Phys. Lett. **25,** 2215 (2008).

[3] H. Takahashi et al., Nature **453,** 376 (2008).

[4] M. Rotter, M. Pangerl, M. Tegel, and D. Johrendt, Angewandte Chemie International Edition **47,** 7949 (2008).

[5] N. Ni, et al., Phys. Rev. B **78,** 014507 (2008).

[6] J.-Q. Yan, et al., Phys. Rev. B **78,** 024516 (2008).

[7] N. Ni, et al., Phys. Rev. B **78,** 014523 (2008).

[8] H. Q. Luo, et al., Supercond. Sci. Technol. **21,** 125014 (2008).

[9] G. F. Chen, et al., Phys. Rev. B **78,** 224512 (2008).

[10] X. F. Wang, et al., Phys. Rev. Lett. **102,** 117005 (2009).

[11] R. Morinaga et al., Jpn. J. Appl. Phys. **48,** 013004 (2009).

[12] Y. Singh, et al., Phys. Rev. B **78,** 104512 (2008).

[13] N. Kumar et al., Phys. Rev. B **79,** 012504 (2009).

[14] P. C. Canfield, et al., Phys. Rev. B **80,** 060501(R) (2009).

[15] H-S Lee, et al., Supercond. Sci. Technol. **22,** 075023(2009).

[16] C. Martin, et al., Phys. Rev. Lett. **102,** 247002 (2009).

[17] L. Fang, et al., J. Cryst. Growth **311,** 358 (2009).

[18] A. Jesche, C. Krellner, M. de Souza, M. Lang, and C. Geibel, arXiv:0909.0903.

[19] C. la Cruz de, et al., Nature **453,** 899 (2008).

[20] T. Nomura, et al., Supercond. Sci. Technol. **21,** 125028 (2008).

[21] M. A. McGuire, et al., Phys. Rev. B **78,** 094517 (2008).

[22] A. Kondrat et al., Eur. Phys. J. B **70**, 461 (2009).

[23] A. Kreyssig, et al., Phys. Rev. B **76**, 054421 (2007).

[24] M. A. Tanatar et al., Phys. Rev. B **79**, 180508 (R) (2009).

[25] A. P. Hammersley, ESRF Internal Report, ESRF97HA02T, "FIT2D: An Introduction and Overview", (1997).